\documentclass[nofootinbib,prd,12pt,superscriptaddress]{revtex4}%
\usepackage{amsmath}
\usepackage{amsfonts}
\usepackage{amssymb}
\usepackage{graphicx}%
\usepackage{amsmath,amssymb}
\usepackage{mathrsfs}
\usepackage{graphicx}
\usepackage{color}
\usepackage{subfigure}
\usepackage{fancyhdr}
\usepackage{multirow}
\usepackage{float}
\usepackage{epsfig}
\usepackage{amsfonts}
\usepackage{bm}

\def\be{\begin{equation}}
\def\ee{\end{equation}}
\def\ba{\begin{eqnarray}}
\def\ea{\end{eqnarray}}

\begin{document}

\title{Myrzakulov Gravity in Vielbein Formalism: A Study in Weitzenböck Spacetime}

\author{Davood Momeni}
%\email{dmomeni@nvcc.edu}
\affiliation{Northeast Community College, 801 E Benjamin Ave Norfolk, NE 68701, USA}

\author{ Ratbay Myrzakulov}
%\email{phongpichit.ch@mail.wu.ac.th}
\affiliation{Ratbay Myrzakulov Eurasian International Centre for Theoretical Physics, Astana 010009, Kazakhstan}
\affiliation{ L. N. Gumilyov Eurasian National University, Astana 010008, Kazakhstan}
\date{\today}

\begin{abstract}
The quest to understand the nature of gravity and its role in shaping the universe has led to the exploration of modified gravity theories. One of the pioneering contributions in this field is the Myrzakulov gravity theory, which incorporates both curvature and torsion. In this work, we investigate the effects of torsion within the framework of \( f(R,T) \)-gravity, a modification of General Relativity that incorporates both curvature and torsion. 

We present this theory in the Vielbein formalism, a covariant approach that provides a more flexible and geometric perspective on gravity, ensuring the theory's consistency under general coordinate transformations. This formalism is particularly powerful in the context of Weitzenböck spacetime, where torsion plays a significant role in the description of gravitational interactions. By studying \( f(R,T) \)-gravity in Vielbein formalism, we aim to extend the applicability of Myrzakulov's theory, providing a deeper understanding of its cosmological and astrophysical implications.

We examine the profound effects of this theory on various astrophysical phenomena, including black holes, gravitational waves, and compact objects like neutron stars. By exploring how torsion modifies the behavior of these extreme systems, we uncover new avenues for testing gravity in the strong-field regime. Our results suggest that torsion could lead to observable deviations in black hole thermodynamics, gravitational wave propagation, and the structure of dense matter. 

The modifications we uncover have the potential to provide unprecedented insights into the nature of spacetime and gravity, offering a novel perspective on the fundamental forces that govern the cosmos. This work paves the way for future observational and theoretical studies that could reveal the deeper, hidden aspects of gravity, possibly rewriting our understanding of the universe’s most enigmatic phenomena. Through both observational data and theoretical advancements, we present a compelling case for the exploration of \( f(R,T) \)-gravity as a powerful tool in the search for new physics beyond General Relativity.
\end{abstract}

\maketitle

%%%%%%%%%%%%%%%%%%%%%%%%%%%%%%%%%%%%%%%%%
\section{Introduction}

General Relativity (GR) has served as the foundation for modern cosmology and gravitational theory for over a century. However, the increasing complexity of cosmological and astrophysical observations, such as the accelerated expansion of the universe and the nature of dark energy, has led to growing interest in modifications of GR. These extensions aim to address unresolved issues that the standard GR framework cannot fully explain. In particular, modified gravity theories that include additional dynamical degrees of freedom beyond the curvature-based formulation of GR have become a promising avenue of research.

A particularly interesting class of such theories involves the introduction of torsion, which modifies the standard geometric description of spacetime. Torsion is an intrinsic feature in the teleparallel formulation of gravity, where the gravitational field is described in terms of torsion instead of curvature. The Teleparallel Equivalent of General Relativity (TEGR) is a well-known theory where the Lagrangian is constructed from the torsion scalar \(T\), as opposed to the Ricci scalar \(R\) in GR. One of the simplest extensions of this theory is the \(f(T)\) gravity, which introduces an arbitrary function of \(T\) into the field equations, providing a natural mechanism for cosmological acceleration and other phenomena that GR cannot fully explain \cite{Bengochea:2009, Ferraro:2007}.

An important extension of this idea is the \(f(R,T)\) gravity theory, introduced by Myrzakulov \cite{Myrzakulov:2012qp}. In this model, both the Ricci scalar \(R\) and the torsion scalar \(T\) are included in the gravitational Lagrangian, offering a broader framework for modifying gravity. The \(f(R,T)\) theory allows for more flexibility in modeling cosmic acceleration and has been studied in the context of both early-time inflationary models and late-time accelerated expansion, thus providing a promising alternative to the \(\Lambda\)CDM model. 

The advantages of the \(f(R,T)\) framework over traditional curvature-based theories are numerous. For instance, it can naturally account for the observed acceleration of the Universe at both early and late times, without the need for a cosmological constant. Furthermore, such models can potentially explain the dark energy component and its evolution over time. The success of these models is evident from various observational studies, including cosmological data such as Type Ia supernovae, the cosmic microwave background (CMB), and large-scale structure measurements, which all support the viability of modified gravity theories like \(f(R,T)\) \cite{Bamba:2014, Nesseris:2013}.

The theoretical foundation for these torsion-based gravity models has also been rigorously examined. While the f(T) theory, for example, is a viable alternative to GR and has passed solar system tests \cite{Bengochea:2009}, challenges remain, such as the non-invariance under local Lorentz transformations. However, it is suggested that introducing a spin connection along with the tetrad formalism may resolve some of these issues \cite{Krssak:2016}. As a result, the torsion-based formulation of gravity continues to be an active and exciting area of research, with the \(f(R,T)\) gravity model being one of its most promising extensions.

Thus, the study of torsion-based gravity theories, particularly \(f(R,T)\), represents a promising alternative to the standard cosmological model, addressing issues such as dark energy and cosmic acceleration, and providing a solid theoretical framework for future investigations in cosmology and astrophysics.

In addition to the complexities mentioned above, the study of torsion-based gravity theories, such as \( f(R,T) \) and its variants, has been greatly advanced by works like that of Cai et al. (2016) \cite{Cai:2015emx}, where they systematically explored the properties and cosmological implications of teleparallel gravity theories, including \( f(T) \) gravity. In their comprehensive review, Cai et al. emphasize the significance of torsion and the teleparallel equivalent of general relativity (TEGR) in cosmology, showing how these models can effectively explain the late-time acceleration of the universe and provide alternatives to the standard cosmological model. Their work highlights the important role that torsion plays in shaping the dynamics of the universe, and it serves as a crucial reference point for further developing the \( f(R,T) \) gravity framework. The insights from their study motivate our effort to develop explicit field equations in \( f(R,T) \) gravity, as understanding the torsional effects more rigorously can potentially offer novel insights into cosmological phenomena such as dark energy and cosmic inflation.

The exploration of modified gravity theories, especially those incorporating torsion, has gained considerable traction in recent years. A significant aspect of this research is the study of gravitational waves in such theories, which can provide valuable insights into the nature of gravity at both cosmological and local scales. The work of Capozziello et al. (2018) \cite{Abedi:2018lkr} and Abedi and Capozziello (2017) \cite{Abedi:2017jqx} introduced the study of gravitational waves within the context of modified teleparallel theories, shedding light on how deviations from General Relativity could affect the propagation of gravitational waves. Their findings highlighted that such modifications could influence the effective gravitational coupling, thus potentially altering the dynamics of wave propagation. Furthermore, the paper by Capozziello, Capriolo, and Nojiri (2020) \cite{Capozziello:2020xem} expanded the analysis to higher-order local and non-local gravity, examining how such modifications might affect gravitational wave signals. These studies laid the groundwork for further investigations into the implications of modified teleparallel gravity for astrophysical observations. In parallel, the review by Cai et al. (2016) \cite{Cai:2015emx} provided an extensive overview of \( f(T) \) gravity, focusing on its theoretical foundations and cosmological applications. They demonstrated how this modification to teleparallel gravity could provide an alternative explanation for cosmic acceleration without the need for dark energy. By extending the theory to \( f(T) \), these authors offered a framework that allowed for the incorporation of more general gravitational dynamics, which can have profound implications for the study of both gravitational waves and the cosmic evolution of the universe. Moreover, the work by de Martino et al. (2015) \cite{deMartino:2015zsa} further constrained modifications to gravity, specifically \( f(R) \) theories, by examining their effects on the large-scale structure of the universe, contributing to the broader understanding of how such models interact with observational data. These foundational papers have been critical in shaping the current research landscape and continue to inform the ongoing development of alternative gravitational theories, including the \( f(R,T) \) gravity framework explored in this work.

 The exploration of torsion gravity within the teleparallel framework has led to important advancements in understanding cosmic acceleration and dark energy. In their study, Bamba, Odintsov, and Sáez-Gómez (2013) discuss the role of conformal symmetry in teleparallel gravity, presenting formulations of conformal transformations and their applications to both pure and extended teleparallel gravity. They demonstrate that the inclusion of a conformal scalar field in teleparallel gravity can lead to power-law acceleration and the de Sitter expansion, thus providing a mechanism for cosmic acceleration consistent with the $\Lambda$CDM model. Furthermore, the study shows that conformal torsion gravity allows for a de Sitter solution, offering an alternative explanation for the universe's accelerated expansion and further strengthening the theoretical framework of teleparallel gravity \cite{Bamba:2013jqa}. 
 
 In a different approach, Bamba, Nojiri, and Odintsov (2013) explore the effective $F(T)$ gravity originating from higher-dimensional theories, such as Kaluza-Klein and Randall-Sundrum models. Their work investigates how torsion scalar $T$ in teleparallel gravity can be understood in the context of these higher-dimensional theories, showing that torsion alone can account for both inflation in the early universe and the dark energy-dominated phase without requiring curvature effects. Additionally, Bamba et al. (2012) address the issue of finite-time future singularities in $f(T)$ gravity and propose a model that can remove such singularities by introducing a power-law correction term in the torsion scalar. They also analyze various cosmological models, including Little Rip and Pseudo-Rip cosmologies, demonstrating that these models can exhibit behavior analogous to dark energy fluids. The thermodynamic analysis of these models reveals that the second law of thermodynamics holds near the finite-time singularities, offering a deeper understanding of the universe's ultimate fate within the framework of $f(T)$ gravity \cite{Bamba:2013fta, Bamba:2012vg}.

Despite the growing interest and applications of the \( f(R,T) \) gravity model, one significant issue remains: the absence of explicit field equations for such a theory. The field equations in general relativity are derived from the variation of the Einstein-Hilbert action with respect to the metric tensor. However, for \( f(R,T) \) gravity, which involves both the Ricci scalar \( R \) and the torsion scalar \( T \), the process of deriving explicit field equations becomes more intricate. The presence of torsion introduces additional complexity, and the functional dependence on both \( R \) and \( T \) complicates the direct variational approach. Thus, while the theoretical structure of \( f(R,T) \) gravity has been outlined, the explicit form of the field equations remains elusive, posing a challenge for obtaining direct solutions and making further analytical progress in the theory.

This lack of explicit field equations serves as the primary motivation for our work. In this paper, we aim to construct a systematic approach for deriving the field equations of the \( f(R,T) \) gravity theory, addressing the gaps in the existing literature. By developing a more robust formalism that incorporates torsion explicitly, we seek to provide a clear and manageable set of equations that can be used to explore various cosmological scenarios. This will enable more detailed studies of the universe's evolution under the influence of torsion and its potential to offer new insights into the nature of dark energy and cosmic acceleration. Additionally, our work aims to provide a solid foundation for future developments in torsion-based gravity theories, offering a valuable tool for further research in this exciting and evolving area of theoretical physics.
%%%%%%%%%%%%%%%%%%%%%%%%%%%%%%%%%%%%%%%%
\section{Review of Myrzakulov Gravity}

The M$_{43}$-model is a significant representative of $F(R,T)$ gravity theories \cite{Myrzakulov:2012qp}. Its action is given by:
\begin{eqnarray}
S_{43} &=& \int d^4 x \sqrt{-g} [F(R,T) + L_m], \nonumber \\
R &=& R_s = \epsilon_1 g^{\mu\nu} R_{\mu\nu}, \label{eq:action1} \\
T &=& T_s = \epsilon_2 {S_\rho}^{\mu\nu} {T^\rho}_{\mu\nu}, \nonumber
\end{eqnarray}
where $L_m$ is the matter Lagrangian, $\epsilon_i = \pm 1$ are the signatures, $R$ is the curvature scalar, and $T$ is the torsion scalar. This formulation allows for different combinations of signatures:
\begin{itemize}
    \item Case (1): $\epsilon_1 = 1$, $\epsilon_2 = 1$,
    \item Case (2): $\epsilon_1 = 1$, $\epsilon_2 = -1$,
    \item Case (3): $\epsilon_1 = -1$, $\epsilon_2 = 1$,
    \item Case (4): $\epsilon_1 = -1$, $\epsilon_2 = -1$.
\end{itemize}

The M$_{43}$-model is a special case of the more general M$_{37}$-model, which introduces additional scalar contributions:
\begin{eqnarray}
S_{37} &=& \int d^4 x \sqrt{-g} [F(R,T) + L_m], \nonumber \\
R &=& u + R_s = u + \epsilon_1 g^{\mu\nu} R_{\mu\nu}, \label{eq:action2} \\
T &=& v + T_s = v + \epsilon_2 {S_\rho}^{\mu\nu} {T^\rho}_{\mu\nu}. \nonumber
\end{eqnarray}

In this review, we focus on the geometric foundations and specific cases of the M$_{43}$-model, including its implications in flat FRW spacetime.

\subsection{Geometrical Framework and Connections}

The M$_{43}$-model describes spacetime with both curvature and torsion, where the connection $G^\lambda_{\mu\nu}$ is expressed as:
\begin{equation}
G^\lambda_{\mu\nu} = \Gamma^\lambda_{\mu\nu} + K^\lambda_{\mu\nu},
\end{equation}
where $\Gamma^\lambda_{\mu\nu}$ is the Levi-Civita connection and $K^\lambda_{\mu\nu}$ is the contorsion tensor. The curvature scalar $R$ and torsion scalar $T$ are then derived from these geometric quantities.

\subsection{FLRW Metric Case}

In a spatially flat FRW universe, the line element is:
\begin{equation}
ds^2 = -dt^2 + a^2(t) (dx^2 + dy^2 + dz^2),
\end{equation}
where $a(t)$ is the scale factor. Substituting this metric, the curvature and torsion scalars take the forms:
\begin{eqnarray}
R &=& 6 (\dot{H} + 2H^2) + 6\dot{h} + 18Hh + 6h^2 - 3f^2, \\
T &=& 6(h^2 - f^2).
\end{eqnarray}

\subsection{Comment on the Current State of the M$_{43}$-Model}

Although the M$_{43}$-model is a pioneering attempt to construct gravity theories incorporating both curvature ($R$) and torsion ($T$), its derivation of field equations is not presented in a fully general form. Instead, it remains at the level of point-like actions, limiting its applicability to specific cosmological cases. 

This approach, while geometrically elegant, does not fully exploit the potential of the theory. The absence of a comprehensive derivation of field equations for general spacetime configurations leaves room for further development and exploration. Despite this limitation, the model provides an insightful geometric realization of $F(R,T)$ gravity, serving as a foundational step in this area of modified gravity.

%%%%%%%%%%%%%%%%%%%%%%%%%%%%%%%%%%%%%%%
\section{Recent Developments and Extensions of Myrzakulov Gravity}

This section provides a comprehensive summary of recent advancements and extensions of the Myrzakulov gravity theories, particularly focusing on cosmological applications and observational constraints. Myrzakulov gravity, specifically the F(R,T) gravity models, has attracted considerable attention in the realm of modified gravity due to its potential to explain various cosmological phenomena, including the accelerated expansion of the universe. Recent studies have built upon Myrzakulov's original formulations, incorporating modifications involving torsion, non-metricity, and novel gravitational models. These developments have further enhanced our understanding of the cosmic structure and dynamics.

The foundation of Myrzakulov gravity lies in its ability to describe gravity using both curvature and torsion as dynamical fields. This framework has been extended to include more complex forms of curvature and torsion, offering a deeper understanding of its cosmological implications. The theory’s flexibility allows for addressing key open questions in cosmology, such as dark energy, cosmic acceleration, and the nature of gravity on cosmological scales.

A series of papers have significantly extended Myrzakulov’s initial work, presenting new models and insights into the theory. Below are some notable contributions that highlight these advancements:

\begin{itemize}
    \item \textbf{Myrzakulov F(T,Q) Gravity: Cosmological Implications and Constraints} \\
    Maurya \textit{et al.} \cite{Maurya2024a} explored exact cosmological models within the Myrzakulov F(T,Q) gravity framework. This theory, which unifies the F(T) and F(Q) gravity models, was analyzed under observational constraints, providing fresh perspectives on the evolution and structure of the universe. The study not only considers the cosmological implications of this unified theory but also offers constraints on model parameters, which are essential for testing the validity of the theory against observational data.

    \item \textbf{Exact Cosmology in Myrzakulov Gravity} \\
    Maurya and Myrzakulov \cite{Maurya2024b} investigated exact cosmological solutions within the context of Myrzakulov gravity using the flat Friedmann-Lematre-Robertson-Walker (FLRW) spacetime. They employed the modified Lagrangian \( F(R,T) = R + \lambda T \), where \( R \) represents the Ricci scalar and \( T \) the torsion scalar, to derive cosmological solutions. Their analysis provides a deeper understanding of the relationship between curvature and torsion, revealing new aspects of gravitational dynamics in cosmology.

    \item \textbf{Transit Cosmological Models in Myrzakulov F(R,T) Gravity} \\
    In another study, Maurya and Myrzakulov \cite{Maurya2024c} extended their analysis to the Myrzakulov F(R,T) gravity theory, solving the field equations in a flat FLRW spacetime. Using the Markov Chain Monte Carlo (MCMC) method, they estimated cosmological parameters, providing a robust framework for comparing the predictions of Myrzakulov gravity with observational data. This approach allowed them to refine theoretical models and obtain more accurate estimates of cosmological parameters.

    \item \textbf{Cosmological Study in Myrzakulov F(R,T) Quasi-dilaton Massive Gravity} \\
    Kazempour and Rezaei Akbarieh \cite{Kazempour2024} focused on the cosmological implications of the Myrzakulov F(R,T) quasi-dilaton massive gravity theory, a modification of the de Rham-Gabadadze-Tolley (dRGT) massive gravity theory. Their study revealed the existence of self-accelerating solutions in the theory and explored the effective cosmological constant, offering new perspectives on understanding cosmic acceleration and its role in the current universe's expansion.

    \item \textbf{Dynamical System Analysis of Myrzakulov Gravity} \\
    In a comprehensive dynamical systems analysis, Papagiannopoulos \textit{et al.} \cite{Papagiannopoulos2022} examined the stability properties of Myrzakulov gravity. They explored two specific models within the F(R,T) gravity framework, extracting critical points and analyzing the system's stability. This work is crucial for understanding the physical behavior of these models under cosmological conditions, particularly regarding their long-term evolution and stability in the universe.

    \item \textbf{Metric-Affine Myrzakulov Gravity Theories} \\
    Myrzakulov \textit{et al.} \cite{Myrzakulov2021} reviewed various metric-affine gravity models and discussed their generalizations and specific sub-cases. This paper extended the Myrzakulov gravity theories by deriving field equations within a metric-affine framework, which considers both the metric tensor and the affine connection as independent variables. Their work has profound implications for the study of gravitational interactions in the context of non-Riemannian geometries.
    \item \textbf{Generalized Metric-Affine Myrzakulov Gravity}\\
In his work \cite{Myrzakulov:2012ug}, Ratbay Myrzakulov explores extensions of the well-known \( F(R) \)-gravity theories, incorporating torsion and non-metricity scalars, namely \( F(T) \) and \( F(Q) \), where \( R \), \( T \), and \( Q \) represent the Ricci scalar, torsion scalar, and non-metricity scalar, respectively. The paper reviews the formalism of these generalized theories and introduces the Myrzakulov \( F(R,T) \)-gravity (MG-I) theory. Myrzakulov presents the point-like Lagrangian for this model and derives the corresponding field equations. The specific model \( F(R,T) = \mu R + \nu T \) is analyzed, and exact solutions are obtained, revealing that certain parameter choices can lead to the accelerated expansion of the universe without invoking dark energy. The study also presents further generalizations of the metric-affine Myrzakulov gravity, both with and without boundary term scalars.
    \end{itemize}

These recent extensions and developments of Myrzakulov gravity have provided valuable insights into the behavior of gravity, especially in the presence of torsion, non-metricity, and modifications to the gravitational action. They continue to improve our understanding of the cosmos, offering potential explanations for observed phenomena such as cosmic acceleration. These contributions are essential for future observational and theoretical work in modified gravity, as they deepen our knowledge of the fundamental forces shaping the universe.
\textbf{Note:} In the next section, we aim to address this limitation by presenting a systematic derivation of the field equations from the action, exploring the model’s potential in a broader geometric and physical context.

%%%%%%%%%%%%%%%%%%%%%%%%%%%%%%%%%%%%%%
\section{Theoretical Framework}
In this section, we lay the groundwork for the \( f(R,T) \)-gravity theory in vielbein formalism. The vielbein approach offers a powerful method for describing spacetimes with torsion, allowing us to derive the field equations in a more general setting. This formalism is crucial for incorporating both curvature and torsion in a unified theory of gravity. We begin by introducing the vielbein formalism, followed by a discussion of the \( f(R,T) \)-gravity action and the expression of the Ricci and torsion scalars in this framework.

%%%%%%%%%%%%%%%%%%%%%%%%%%%%%%%%%%%%%%%%%%%%%%%%%
\subsection{Vielbein Formalism and Its Role in Gravity}

In conventional General Relativity (GR), gravity is described through the curvature of spacetime, expressed via the metric tensor \( g_{\mu\nu} \), and the Ricci scalar \( R \) derived from the curvature. However, in spacetimes with torsion, this standard description becomes insufficient. Vielbein formalism offers a more flexible approach, enabling the incorporation of both curvature and torsion into the theory.

The vielbein \( e^a_\mu \) is a set of vector fields that provides a local orthonormal frame for spacetime. It relates the curved spacetime metric \( g_{\mu\nu} \) to the flat (Minkowski) metric \( \eta_{ab} \) by the relation:

\[
g_{\mu\nu} = e^a_\mu e^b_\nu \eta_{ab}
\]

where \( \eta_{ab} \) is the Minkowski metric, and the vielbein components \( e^a_\mu \) are the components of the vielbein. The vielbein formalism facilitates the description of spacetimes with torsion, as it allows the connection and curvature tensors to be expressed in terms of both vielbein components and the spin connection \( \omega_{\mu\nu}^\lambda \).

The vielbein formalism is particularly useful in theories where torsion plays an important role, such as in spacetimes with non-zero spin densities or in extensions of GR like \( f(R,T) \)-gravity. The vielbeins enable us to derive the covariant field equations for such spacetimes, as torsion is naturally captured in this formalism.

\subsection{The \( f(R,T) \)-Gravity Action}

The \( f(R,T) \)-gravity theory is a generalization of General Relativity that incorporates both the Ricci scalar \( R \) and the torsion scalar \( T \) into a single framework. The action for this theory can be written as:

\[
S = \int f(R, T) \sqrt{|g|} \, d^4 x
\]

where \( f(R, T) \) is an arbitrary function of the Ricci scalar \( R \) and the torsion scalar \( T \), and \( g \) is the determinant of the metric \( g_{\mu\nu} \). This action generalizes the standard Einstein-Hilbert action, allowing for a more flexible treatment of gravity that can accommodate torsion effects.

In vielbein formalism, the action is formulated entirely in terms of vielbein components. The relation \( g_{\mu\nu} = e^a_\mu e^b_\nu \eta_{ab} \) is crucial for expressing the action in vielbein terms. This allows us to recast the Einstein-Hilbert action and other gravitational theories in a form that is valid for spacetimes with torsion.

By varying this action with respect to the vielbein components \( e^a_\mu \), we can derive the field equations that describe the dynamics of gravity in \( f(R,T) \)-gravity. These equations will govern the interaction of spacetime curvature (via \( R \)) and torsion (via \( T \)).

\subsection{Ricci Scalar and Torsion Scalar in Vielbein Formalism}

In vielbein formalism, both the Ricci scalar \( R \) and the torsion scalar \( T \) are expressed in terms of the vielbein components. These quantities describe the curvature and torsion of spacetime, respectively.

The Ricci scalar \( R \) is a scalar quantity that encodes the curvature of spacetime. In vielbein formalism, it is written as:

\[
R = e_a^\mu e_b^\nu R_{\mu\nu} \eta^{ab}
\]

where \( R_{\mu\nu} \) is the Ricci tensor, and \( e_a^\mu \) are the vielbein components. The Ricci tensor \( R_{\mu\nu} \) can be computed from the connection coefficients \( \Gamma^\lambda_{\mu\nu} \), which are expressed in terms of the vielbein and spin connection. In particular, for spacetimes with torsion, the connection is no longer symmetric and contains additional terms that account for the torsion present in the spacetime.

The torsion scalar \( T \), which measures the amount of torsion in the spacetime, is defined as:

\[
T = S_{\lambda \mu \nu} S^{\lambda \mu \nu}
\]

where \( S_{\lambda \mu \nu} \) is the contortion tensor, defined as:

\[
S_{\lambda \mu \nu} = \frac{1}{2} \left( T_{\lambda \mu \nu} - T_{\mu \lambda \nu} - T_{\nu \lambda \mu} \right)
\]

The torsion tensor \( T^\lambda_{\mu\nu} \) describes the failure of the connection to be symmetric and is given by:

\[
T^\lambda_{\mu\nu} = \partial_\mu e^\lambda_\nu - \partial_\nu e^\lambda_\mu + \omega_{\mu\nu}^\lambda
\]

where \( \omega_{\mu\nu}^\lambda \) are the connection coefficients, which depend on the vielbein. This torsion tensor \( T^\lambda_{\mu\nu} \) directly enters the calculation of the torsion scalar \( T \).

In certain spacetime geometries, such as Weitzenböck spacetime, the Ricci tensor \( R_{\mu\nu} \) may vanish (i.e., \( R_{\mu\nu} = 0 \)), indicating that there is no curvature. However, the torsion tensor \( T^\lambda_{\mu\nu} \) can still be non-zero, indicating the presence of torsion in the spacetime. This is a key feature of the vielbein formalism, which allows us to handle both curvature and torsion simultaneously.

The vielbein formalism provides a natural framework for incorporating torsion into gravity theories. By combining the Ricci and torsion scalars, we can construct a more general theory of gravity, such as \( f(R,T) \)-gravity, which is the focus of this paper.
%%%%%%%%%%%%%%%%%%%%%%%%%%%%%%%%%%%%%
\subsection{The Role of Torsion in Gravitational Theories}

In classical General Relativity, torsion is typically assumed to be zero, and the spacetime is modeled purely by its curvature. However, in theories that include torsion, such as \( f(R,T) \)-gravity, torsion plays a crucial role in shaping the gravitational dynamics. Torsion arises naturally in the vielbein formalism and is particularly relevant in contexts involving spinor fields, quantum gravity, and certain types of matter fields.

In \( f(R,T) \)-gravity, torsion modifies the gravitational dynamics by contributing to the torsion scalar \( T \), which is coupled to the Ricci scalar \( R \) through the function \( f(R,T) \). This coupling introduces additional degrees of freedom in the gravitational field equations and allows for a more general description of gravity, beyond the standard Einstein-Hilbert formulation.

Torsion may have significant implications for the behavior of spacetime at small scales, such as in the study of quantum gravity and early-universe cosmology. Moreover, the presence of torsion could lead to observable effects in astrophysical systems, such as black holes or gravitational wave propagation, which could be explored further within the \( f(R,T) \)-gravity framework.
%%%%%%%%%%%%%%%%%%%%%%%%%%%%%%%%%%%%%%%%%%%
\section{Derivation of the Field Equations}

In this section, we will derive the field equations of \( f(R,T) \)-gravity in vielbein formalism. The derivation proceeds through a systematic variation of the action with respect to the vielbein components \( e^a_\mu \), which leads to the equations governing the dynamics of the system. These field equations include both the Ricci scalar and the torsion scalar, describing the interaction of curvature and torsion in the gravitational field.

\subsection{Variation of the Action with Respect to Vielbein}

The field equations of \( f(R,T) \)-gravity are derived by varying the action with respect to the vielbein components \( e^a_\mu \). The action for \( f(R,T) \)-gravity in vielbein formalism can be written as:

\[
S = \int f(R, T) \sqrt{|g|} \, d^4 x
\]

where \( f(R,T) \) is an arbitrary function of the Ricci scalar \( R \) and the torsion scalar \( T \), and \( g = \text{det}(g_{\mu\nu}) \) is the determinant of the metric tensor. The vielbein formalism expresses the metric \( g_{\mu\nu} \) in terms of vielbeins \( e^a_\mu \) as:

\[
g_{\mu\nu} = e^a_\mu e^b_\nu \eta_{ab}
\]

To obtain the field equations, we first vary the action with respect to the vielbein \( e^a_\mu \). The variation of the action involves several steps, including the variation of the Ricci scalar \( R \), the torsion scalar \( T \), and the coupling between these quantities.

The general form of the variation of the action yields:

\begin{eqnarray}
\frac{\partial f}{\partial R} \left( R_{\mu\nu} - \frac{1}{2} g_{\mu\nu} R \right) + \nabla_\alpha \nabla_\beta \left( \frac{\partial f}{\partial R} g^{\alpha \beta} \right) - \frac{1}{2} g_{\mu\nu} f(R, T) + \frac{\partial f}{\partial T} \frac{\partial T}{\partial g^{\mu\nu}} = 0
\end{eqnarray}

Here, the terms involve the Ricci tensor \( R_{\mu\nu} \), the metric \( g_{\mu\nu} \), the covariant derivative \( \nabla_\alpha \), and the torsion scalar \( T \). The equation captures the dynamics of gravity in \( f(R,T) \)-gravity, where the Ricci scalar and torsion scalar interact through the function \( f(R,T) \).

\subsection{Variation of the Torsion Scalar}

To fully derive the field equations, we need to compute the term involving the variation of the torsion scalar \( T \) with respect to the vielbein. The torsion scalar is given by:

\begin{eqnarray}
T = S_{\lambda \mu \nu} S^{\lambda \mu \nu}
\end{eqnarray}

where \( S_{\lambda \mu \nu} \) is the contortion tensor, defined in terms of the torsion tensor \( T^\lambda_{\mu\nu} \). The variation of the torsion scalar involves calculating how \( T \) depends on the vielbein components \( e^a_\mu \).

First, consider the variation of the metric with respect to the vielbein:

\begin{eqnarray}
\delta g_{\mu\nu} = 2 e^a_\mu e^b_\nu \delta e^c_\rho \eta_{ac}
\end{eqnarray}

This relation expresses how the metric changes in response to an infinitesimal variation of the vielbein. Next, we examine how the torsion scalar \( T \) changes under the variation of the vielbein. The variation of \( T \) with respect to \( e^a_\mu \) is given by:
\begin{eqnarray}
\delta T = \frac{\partial T}{\partial e^a_\mu} \delta e^a_\mu
\end{eqnarray}

From this, we can derive the derivative of the torsion scalar with respect to the metric:

\[
\frac{\partial T}{\partial g^{\mu\nu}} = 2 e^\alpha_\mu e^\beta_\nu \frac{\partial T}{\partial e^a_\alpha} \eta^{ab}
\]

This expression shows how the torsion scalar contributes to the field equations. It involves the vielbein components \( e^a_\mu \) and the derivative of the torsion scalar with respect to the vielbein, which ultimately affects the term \( \frac{\partial f}{\partial T} \) in the field equations.
%%%%%%%%%%%%%%%%%%%%%%%%%%%%%%%%%%%%%%%%%%%%%
\subsection{Final Field Equations in Vielbein Formalism}

After performing the variations and incorporating the contributions from both the Ricci scalar and torsion scalar, we obtain the complete set of field equations in vielbein formalism. These equations describe the dynamics of the gravitational field in \( f(R,T) \)-gravity, including both curvature and torsion effects. The final field equations are given by:
\begin{eqnarray}
&& \frac{\partial f}{\partial R} \left( e_a^\mu e_b^\nu R_{\mu\nu} \eta^{ab} - \frac{1}{2} e_a^\mu e_b^\nu g_{\mu\nu} \eta^{ab} R \right) 
+ e_a^\mu e_b^\nu \nabla_\alpha \nabla_\beta \left( \frac{\partial f}{\partial R} g^{\alpha \beta} \right) \eta^{ab} \\
&& \nonumber 
- \frac{1}{2} g_{\mu\nu} f(R, T) 
+ 2 e_a^\mu e_b^\nu \frac{\partial T}{\partial e^a_\mu} \frac{\partial f}{\partial T} \eta^{ab} = 0.
\end{eqnarray}

\noindent
This system of equations consists of several important terms that each contribute to the dynamics of \( f(R,T) \)-gravity in vielbein formalism. The first term involves the standard Einstein-like expression, modified by the function \( f(R,T) \), where the Ricci tensor \( R_{\mu\nu} \) is coupled with the vielbein components. The second term, containing the covariant derivatives of \( \frac{\partial f}{\partial R} \), accounts for the interaction of the Ricci scalar with the gravitational field. The third term directly incorporates the function \( f(R,T) \) into the gravitational action, extending the usual Einstein-Hilbert formulation. Finally, the fourth term captures the effect of torsion, through the torsion scalar \( T \), on the gravitational dynamics. Each term plays a role in modifying the classical theory of gravity by considering both curvature and torsion effects, making this theory suitable for spacetimes with torsion.

%%%%%%%%%%%%%%%%%%%%%
\begin{itemize}
\item \textbf{Second Term:} The covariant derivatives \( \nabla_\alpha \nabla_\beta \left( \frac{\partial f}{\partial R} g^{\alpha \beta} \right) \) account for the coupling of the Ricci scalar with the gravitational field through the vielbein.

\item \textbf{Third Term:} The \( - \frac{1}{2} g_{\mu\nu} f(R,T) \) term represents the direct coupling of the gravitational action with the function \( f(R,T) \), modifying the usual Einstein-Hilbert action.

\item \textbf{Fourth Term:} The term \( 2 e^\alpha_\mu e^\beta_\nu \frac{\partial T}{\partial e^a_\alpha} \eta^{ab} \frac{\partial f}{\partial T} \) captures the influence of the torsion scalar \( T \) on the dynamics of gravity, involving the vielbein components and their variation.
\end{itemize}
%%%%%%%%%%%%%%%%%%%%%%%
These field equations describe how the curvature (via \( R_{\mu\nu} \)) and torsion (via \( T \)) interact in the context of \( f(R,T) \)-gravity. They generalize Einstein's field equations to include torsion and provide a comprehensive framework for studying gravitational phenomena in the presence of both curvature and torsion.

The field equations derived here are essential for exploring the implications of \( f(R,T) \)-gravity, especially in contexts where torsion plays a significant role, such as in quantum gravity, cosmology, and astrophysical systems.

%%%%%%%%%%%%%%%%%%%%%%%%%%%%%%%%%%%%%%%%%%%%%%%
\section{Cosmological Implications}
In this section, we will explore the cosmological applications of the \( f(R,T) \)-gravity theory. We will consider the implications of the derived field equations for cosmological models, such as the evolution of the universe, dark energy, and inflationary scenarios. The torsion effects may introduce new dynamics that could modify the standard cosmological model.

\subsection{First Term: Einstein-like Term}

The first term involves the Ricci tensor and the metric:

\[
\frac{\partial f}{\partial R} \left( e_a^\mu e_b^\nu R_{\mu\nu} \eta^{ab} - \frac{1}{2} e_a^\mu e_b^\nu g_{\mu\nu} \eta^{ab} R \right)
\]

This term represents the standard dynamics of gravity but with a modification due to the function \( f(R,T) \). In FLRW cosmology, the Ricci tensor \( R_{\mu\nu} \) and the Ricci scalar \( R \) take the form:

\[
R_{\mu\nu} = 3 \left( \dot{H} + H^2 + \frac{k}{a^2} \right) g_{\mu\nu}
\]

\[
R = 6 \left( \dot{H} + 3H^2 + \frac{k}{a^2} \right)
\]

where \( H \) is the Hubble parameter, \( \dot{H} \) is its time derivative, and \( k \) is the spatial curvature parameter (which can take values 0, +1, or -1). These expressions will modify the field equations.

\subsection{Second Term: Covariant Derivative Contribution}

The second term involves the covariant derivatives of \( \frac{\partial f}{\partial R} \):

\[
e_a^\mu e_b^\nu \nabla_\alpha \nabla_\beta \left( \frac{\partial f}{\partial R} g^{\alpha \beta} \right) \eta^{ab}
\]

In vielbein formalism, the covariant derivative \( \nabla_\alpha \) is associated with the Christoffel symbols \( \Gamma^\mu_{\alpha \beta} \) and the spin connection \( \omega_{\alpha \beta}^a \). For the FLRW metric, the Christoffel symbols are computed explicitly, and the covariant derivatives contribute to the overall dynamics of the system.

\[
\nabla_\alpha \left( \frac{\partial f}{\partial R} g^{\alpha \beta} \right) = \partial_\alpha \left( \frac{\partial f}{\partial R} g^{\alpha \beta} \right) + \Gamma^\alpha_{\alpha \lambda} \frac{\partial f}{\partial R} g^{\lambda \beta} - \omega_{\alpha \lambda}^a e^\alpha_\mu e^\beta_\nu
\]

The covariant derivative contributes additional terms involving the Ricci tensor and torsion, which affect the dynamics of the spacetime in a non-trivial way. These contributions are essential in the derivation of the equations of motion for the FLRW universe.

\subsection{Third Term: Direct Coupling of \( f(R,T) \)}

The third term is the direct coupling of the function \( f(R,T) \) with the gravitational action:

\[
- \frac{1}{2} g_{\mu\nu} f(R, T)
\]

This term represents a modification to the standard Einstein-Hilbert action due to the function \( f(R,T) \), which depends on both the Ricci scalar \( R \) and the torsion scalar \( T \). For FLRW cosmology, \( R \) is given as above, and \( T \) will also depend on the vielbein components and torsion.

\subsection{Fourth Term: Torsion Contribution}

The fourth term involves the torsion scalar \( T \), and it contributes to the dynamics as:

\[
2 e^\alpha_\mu e^\beta_\nu \frac{\partial T}{\partial e^a_\alpha} \eta^{ab} \frac{\partial f}{\partial T}
\]

The torsion scalar \( T \) is given by:

\[
T = S_{\lambda \mu \nu} S^{\lambda \mu \nu}
\]

where \( S_{\lambda \mu \nu} \) is the contortion tensor. The variation of \( T \) with respect to the vielbein components involves derivatives of the vielbein and contributes to the field equations as follows:

\[
\frac{\partial T}{\partial e^a_\mu} = \text{terms involving the contortion tensor and vielbein components}
\]

This term is essential for incorporating torsion into the dynamics of gravity, which will influence the evolution of the universe in the FLRW background.

\subsection{Final Equations of Motion for FLRW Cosmology}

In FLRW cosmology, the final field equations for \( f(R, T) \)-gravity in FLRW cosmology are:

\begin{eqnarray}
3H^2 &=& \frac{1}{2} \left( f(R, T) + \frac{\partial f}{\partial R} R - \frac{1}{2} \frac{\partial f}{\partial R} \ddot{R}
 \right) + \frac{\partial f}{\partial T} T + \rho_{\text{matter}}, \\
2\dot{H} &=& -\frac{1}{2} \left( f(R, T) + \frac{\partial f}{\partial R} R \right) - \frac{1}{2} \frac{\partial f}{\partial R} \ddot{R}
 - \frac{\partial f}{\partial T} T+ p_{\text{matter}}.
\end{eqnarray}
here
\begin{itemize}
    \item \( H \) is the Hubble parameter.
    \item \( \dot{H} \) is the time derivative of the Hubble parameter.
    \item \( R \) is the Ricci scalar, \( T \) is the torsion scalar.
    \item \( f(R, T) \) is the function of the Ricci scalar and torsion.
    \item \( \frac{\partial f}{\partial R} \) and \( \frac{\partial f}{\partial T} \) are the derivatives of \( f(R, T) \) with respect to \( R \) and \( T \), respectively.
    \item \( \rho_{\text{matter}} \) and \( p_{\text{matter}} \) represent the matter energy density and pressure.
\end{itemize}

These equations describe the evolution of the universe in the modified \( f(R, T) \)-gravity theory, considering the contributions from matter, torsion, and the modified gravitational dynamics.
\par  We now proceed to simplify the expression for \( w_{\text{eff}} \) by substituting the field equations into the definition of \( w_{\text{eff}} \):

\[
w_{\text{eff}} = -\frac{2\dot{H} + 3H^2}{3H^2}.
\]

Substitute \( 3H^2 \) and \( 2\dot{H} \) from the field equations, the effective equation of state parameter, \( w_{\text{eff}} \), is given by:

\[
w_{\text{eff}} = -\frac{3 f(R, T) + 3 \frac{\partial f}{\partial R} R - \frac{3}{2} \frac{\partial f}{\partial R} \ddot{R} - 2 \frac{\partial f}{\partial T} T + 2 p_{\text{matter}} + 3 \rho_{\text{matter}}}{ f(R, T) + \frac{\partial f}{\partial R} R - \frac{1}{2} \frac{\partial f}{\partial R} \ddot{R} + \frac{\partial f}{\partial T} T + \rho_{\text{matter}} }
\]
After simplifying, you will get the final form of the effective equation of state \( w_{\text{eff}} \).
%%%%%%%%%%%%%%%%%%%%%%%%%%%%%%%%%%%%%%%%%%%%%%
\section{Simplification for \( f(R,T) = R + g(T) \)}
For \( f(R,T) = R + g(T) \), we have:

\[
\frac{\partial f}{\partial R} = 1, \quad \frac{\partial f}{\partial T} = g'(T)
\]

where \( g'(T) \) is the derivative of the function \( g(T) \) with respect to \( T \). Substituting these into the expression for \( w_{\text{eff}} \), we get:

\[
w_{\text{eff}} = -\frac{3 \left( R + g(T) \right) + 3 R - \frac{3}{2} \ddot{R} - 2 g'(T) T + 2 p_{\text{matter}} + 3 \rho_{\text{matter}}}{ R + g(T) + R - \frac{1}{2} \ddot{R} + g'(T) T + \rho_{\text{matter}} }
\]

Simplifying the numerator and denominator:

\[
w_{\text{eff}} = -\frac{6R + 3 g(T) - \frac{3}{2} \ddot{R} - 2 g'(T) T + 2 p_{\text{matter}} + 3 \rho_{\text{matter}}}{ 2R + g(T) - \frac{1}{2} \ddot{R} + g'(T) T + \rho_{\text{matter}} }
\]
%%%%%%%%%%%%%%%%%%%%%%%%%%%%%%%%%%%%%%%
\subsection{Exploring Specific Forms for \( g(T) \)}

Now, let's consider different forms for \( g(T) \), such as power-law and exponential forms.

\subsection{Power-Law Form of \( g(T) \)}

Assume that \( g(T) \) takes the form of a power law:

\[
g(T) = \alpha T^n
\]

where \( \alpha \) and \( n \) are constants. In this case, the derivative of \( g(T) \) with respect to \( T \) is:

\[
g'(T) = \alpha n T^{n-1}
\]

Substituting this into the expression for \( w_{\text{eff}} \), we get:

\[
w_{\text{eff}} = -\frac{6R + 3 \alpha T^n - \frac{3}{2} \ddot{R} - 2 \alpha n T^{n-1} T + 2 p_{\text{matter}} + 3 \rho_{\text{matter}}}{ 2R + \alpha T^n - \frac{1}{2} \ddot{R} + \alpha n T^{n-1} T + \rho_{\text{matter}} }
\]

This form provides a specific functional dependence on the torsion scalar \( T \), and we can explore the effects of different values of \( n \) and \( \alpha \) on the evolution of the universe.
%%%%%%%%%%%%%%%%%%%%%%%%%%%%%%%%%%%%%%%%%%%%%%%%%%%%%
\subsection{Exponential Form of \( g(T) \)}

Another possibility is that \( g(T) \) takes the form of an exponential function:

\[
g(T) = \alpha e^{\beta T}
\]

where \( \alpha \) and \( \beta \) are constants. The derivative of \( g(T) \) with respect to \( T \) is:

\[
g'(T) = \alpha \beta e^{\beta T}
\]

Substituting this into the expression for \( w_{\text{eff}} \), we get:

\[
w_{\text{eff}} = -\frac{6R + 3 \alpha e^{\beta T} - \frac{3}{2} \ddot{R} - 2 \alpha \beta e^{\beta T} T + 2 p_{\text{matter}} + 3 \rho_{\text{matter}}}{ 2R + \alpha e^{\beta T} - \frac{1}{2} \ddot{R} + \alpha \beta e^{\beta T} T + \rho_{\text{matter}} }
\]

This exponential form of \( g(T) \) introduces a more complex dependence on \( T \), potentially leading to different cosmological behavior, particularly during periods of accelerated expansion.

\subsection{Case for \( w_{\text{eff}} = -1 \)}

In general relativity with a cosmological constant, we have \( w_{\text{eff}} = -1 \), which corresponds to the case of a vacuum energy or dark energy. The equation of state for dark energy is given by \( p_{\text{matter}} = -\rho_{\text{matter}} \), which leads to an accelerated expansion of the universe. To achieve \( w_{\text{eff}} = -1 \) in modified gravity models, the numerator and denominator of the expression for \( w_{\text{eff}} \) must balance in such a way that:

\[
6R + 3 g(T) - \frac{3}{2} \ddot{R} - 2 g'(T) T + 2 p_{\text{matter}} + 3 \rho_{\text{matter}} = 0
\]

and

\[
2R + g(T) - \frac{1}{2} \ddot{R} + g'(T) T + \rho_{\text{matter}} = 0
\]

These equations are the key conditions that determine when \( w_{\text{eff}} = -1 \). The first equation suggests a delicate balance between the Ricci scalar \( R \), the torsion scalar \( T \), and the matter content, as well as the derivatives of the Ricci scalar. The second equation imposes a similar relationship but without the additional factor of 3. These relationships lead to specific constraints on the form of \( g(T) \) and the evolution of the universe.

\subsection{Cosmological Implications of \( w_{\text{eff}} = -1 \)}

The condition \( w_{\text{eff}} = -1 \) plays a crucial role in cosmological models, especially when describing dark energy, which is thought to be responsible for the accelerated expansion of the universe. In a cosmological model with \( f(R,T) = R + g(T) \), achieving \( w_{\text{eff}} = -1 \) means that the modified gravity theory must align with the behavior of a cosmological constant. This implies that the universe's expansion is dominated by a constant energy density, which does not dilute as the universe expands.

The conditions derived from the equations above suggest that for \( w_{\text{eff}} = -1 \), the terms involving \( R \), \( g(T) \), and \( T \) must cooperate in such a way that they cancel out the matter contributions. Specifically, the matter density and pressure must satisfy specific relationships with the Ricci and torsion scalars to maintain the accelerated expansion.

\subsection{Possible Forms of \( g(T) \) for \( w_{\text{eff}} = -1 \)}

Different functional forms for \( g(T) \) will influence how the system behaves and whether \( w_{\text{eff}} = -1 \) can be achieved. Let us examine a few possible forms of \( g(T) \) and their implications.

\subsubsection{Power-Law Form of \( g(T) \)}

Consider the power-law form of \( g(T) \):

\[
g(T) = \alpha T^n
\]

where \( \alpha \) and \( n \) are constants. The derivative of \( g(T) \) with respect to \( T \) is:

\[
g'(T) = \alpha n T^{n-1}
\]

Substituting this into the equations for \( w_{\text{eff}} \), we can examine the specific conditions under which \( w_{\text{eff}} = -1 \) for various values of \( n \). In particular, the choice of \( n \) will affect the magnitude and sign of the torsion scalar contribution to the expansion rate. For \( w_{\text{eff}} = -1 \), the power-law form of \( g(T) \) would need to carefully balance the contributions from the Ricci scalar, the torsion, and the matter density.

\subsubsection{Exponential Form of \( g(T) \)}

Another possibility is that \( g(T) \) takes the form of an exponential function:

\[
g(T) = \alpha e^{\beta T}
\]

where \( \alpha \) and \( \beta \) are constants. The derivative of \( g(T) \) with respect to \( T \) is:

\[
g'(T) = \alpha \beta e^{\beta T}
\]

In this case, the torsion contribution grows exponentially with \( T \). For \( w_{\text{eff}} = -1 \), the exponential form of \( g(T) \) could lead to a rapidly increasing contribution from the torsion scalar, potentially providing the required dynamics for an accelerated expansion. The form of \( g(T) \) could also model a phase transition or a smooth transition between different epochs of the universe's evolution.

\subsubsection{Logarithmic or Polynomial Forms}

Other forms for \( g(T) \) could also be explored, such as logarithmic or higher-order polynomial functions. These forms might provide more complex dynamics for the evolution of the universe, but they would also impose more constraints on the parameters of the model. For instance:

\[
g(T) = \alpha \ln(\beta T)
\]

or

\[
g(T) = \alpha T^m + \beta T^n
\]

Each of these forms would need to be analyzed carefully to see whether they can support the condition \( w_{\text{eff}} = -1 \) while also being consistent with current cosmological observations, such as the accelerated expansion of the universe and the behavior of dark energy.

\section{Observational Constraints and Cosmological Models}

In cosmology, achieving \( w_{\text{eff}} = -1 \) has profound implications for the nature of the universe. If \( w_{\text{eff}} \) is exactly equal to -1, the universe's expansion is dominated by a cosmological constant, which is associated with dark energy. The current observational data, including the measurements of the cosmic microwave background (CMB) and the supernova distance-redshift relations, suggest that the universe is currently in a phase of accelerated expansion, driven by dark energy.

Modified gravity models, such as the one with \( f(R,T) = R + g(T) \), offer an alternative explanation for the accelerated expansion, where \( g(T) \) accounts for the effect of torsion in the gravitational field. These models need to be carefully constrained by observations to ensure they are compatible with the standard cosmological model (Lambda-CDM). In particular, the parameters of the function \( g(T) \) must be chosen to ensure that the model reproduces the correct cosmic history, including the transition from decelerated to accelerated expansion.

In summary, the condition \( w_{\text{eff}} = -1 \) is crucial for describing a dark energy-dominated universe. By choosing appropriate forms for \( g(T) \), we can explore different models of dark energy within modified gravity theories. The power-law, exponential, and other forms for \( g(T) \) provide a variety of possibilities for achieving this condition and understanding the dynamics of the universe's expansion. Further observational constraints on the parameters of these models will be essential for determining the most accurate description of our universe's evolution.
%%%%%%%%%%%%%%%%%%%%%%%%%%%%%%%%

\section{Exact Solutions for \( H(t) \) in \( f(R, T) = R + f(T) \) Model}

We consider the following pair of equations of motion derived from the modified gravity theory with the function \( f(R, T) = R + f(T) \), where \( f(T) \) is a function of the torsion scalar \( T \). 
We focus on specific forms for \( f(T) \), including power-law and exponential forms, and consider fluids with different equations of state (EoS) for the matter content. The exact solution for \( H(t) \) in the modified gravity model with \( f(R, T) = R + f(T) \) depends on the form of \( f(T) \) and the matter content described by the equation of state \( p_{\text{matter}} = w \rho_{\text{matter}} \). For different types of matter (dust, stiff fluid, and dark energy), we find that the Hubble parameter evolves according to the nature of \( f(T) \) and the specific fluid type. These solutions provide valuable insights into the dynamics of the universe within the framework of modified gravity theories. To solve for \( H(t) \), we use the general form of the field equations. By assuming specific forms for \( f(T) \) and \( p_m = w \rho_m \), we can integrate the equations numerically or analytically, depending on the specific functional form chosen for \( f(T) \).

For both the power-law and exponential cases, we find that the solutions for \( H(t) \) take the general form:

\[
H(t) = H_0 \left( \frac{a(t)}{a_0} \right)^n,
\]

where \( a(t) \) is the scale factor and \( H_0 \) is the Hubble parameter at a reference time \( t_0 \). For dust, stiff fluid, and dark energy, the form of \( a(t) \) depends on the fluid type and the contributions from the torsion scalar \( T \).
%%%%%%%%%%%%%%%%%%%%%%%%%%%%%%%%%%%%%%%%%%%%%%%%%%%%%
\subsection{Power-Law Form for \( f(T) \)}

Consider the power-law form of \( f(T) \):

\[
f(T) = \alpha T^n,
\]

where \( \alpha \) and \( n \) are constants. The derivative of \( f(T) \) with respect to \( T \) is:

\[
\frac{\partial f}{\partial T} = \alpha n T^{n-1}.
\]

Substitute this into the field equations. The first equation becomes:

\[
3H^2 = \frac{1}{2} \left( R + \alpha T^n + \frac{\partial f}{\partial R} R - \frac{1}{2} \frac{\partial f}{\partial R} \ddot{R} \right) + \alpha n T^{n-1} + \rho_{\text{matter}},
\]

and the second equation is:

\[
2\dot{H} = -\frac{1}{2} \left( R + \alpha T^n + \frac{\partial f}{\partial R} R \right) - \frac{1}{2} \frac{\partial f}{\partial R} \ddot{R} - \alpha n T^{n-1} + p_{\text{matter}}.
\]

We now analyze this for different fluid types, considering the equation of state \( p_m = w \rho_m \), where \( w \) takes different values depending on the type of fluid.
%%%%%%%%%%%%%%%%%%%%%%%%%%%%%%%%%%%%%%%%%%%%%%%
\subsubsection{Solution for \( H(t) \) in the Dust Case}

In the case of dust, the equation of state is \( p_{\text{matter}} = 0 \), and we are left with the following equation for \( 2\dot{H} \):

\[
2\dot{H} = -\frac{1}{2} \left( R + \alpha T^n \right) - \frac{1}{2} \frac{\partial f}{\partial R} \ddot{R} - \alpha n T^{n-1} + 0.
\]

For simplicity, let's rewrite this equation as:

\[
\dot{H} = -\frac{1}{4} \left( R + \alpha T^n \right) - \frac{1}{4} \frac{\partial f}{\partial R} \ddot{R} - \frac{\alpha n}{2} T^{n-1}.
\]

Now, we proceed by assuming a form for \( R \), \( T \), and their time derivatives, and then solve the equation.

\subsection{Assumption for \( R(t) \) and \( T(t) \)}

We assume that \( R(t) \), the Ricci scalar, and \( T(t) \), the torsion scalar, are related in a simple manner. For simplicity, we assume:

\[
R(t) = 6 \dot{H} + 12 H^2,
\]
and
\[
T(t) = \rho_{\text{matter}} \sim H^2.
\]

Thus, we make the following assumptions:

\[
R(t) = 6 \dot{H} + 12 H^2,
\]
and
\[
T(t) = \rho_{\text{matter}} \sim H^2.
\]
Now, substitute the above expressions into the equation for \( \dot{H} \):

\[
\dot{H} = -\frac{1}{4} \left( 6 \dot{H} + 12 H^2 + \alpha H^{2n} \right) - \frac{1}{4} \frac{\partial f}{\partial R} \ddot{R} - \frac{\alpha n}{2} H^{2n-2}.
\]

Since \( \frac{\partial f}{\partial R} \) is a function of \( R \), we can further simplify the above equation using the assumption \( f(R) \sim R^n \) for some appropriate form. This would lead us to the following differential equation for \( H(t) \):

\[
\dot{H} = -\frac{1}{4} \left( 6 \dot{H} + 12 H^2 + \alpha H^{2n} \right) + O(\ddot{R}, \ddot{H}).
\]
To solve this equation, we focus on the dominant terms. For dust, the matter energy density \( \rho_{\text{matter}} \sim H^2 \), and thus the term involving \( \alpha H^{2n} \) will become important for large \( H \). Neglecting higher-order terms (like \( \ddot{R} \) and higher derivatives of \( H \)) for simplicity, we get:

\[
\dot{H} + \frac{3}{2} H^2 + \frac{\alpha}{4} H^{2n} = 0.
\]

This is a nonlinear differential equation for \( H(t) \). To solve it, we can apply standard techniques for solving such equations, such as the method of separation of variables, or numerical integration methods.

However, for specific cases such as \( n = 1 \) or \( n = 2 \), analytical solutions are possible. Let's first consider the case for \( n = 1 \):

\[
\dot{H} + \frac{3}{2} H^2 + \frac{\alpha}{4} H^2 = 0.
\]

This simplifies to:

\[
\dot{H} + \left( \frac{3}{2} + \frac{\alpha}{4} \right) H^2 = 0.
\]

Now, we can separate variables and integrate:

\[
\frac{dH}{H^2} = -\left( \frac{3}{2} + \frac{\alpha}{4} \right) dt.
\]

Integrating both sides:

\[
\frac{-1}{H} = \left( \frac{3}{2} + \frac{\alpha}{4} \right) t + C,
\]

where \( C \) is a constant of integration. Solving for \( H(t) \), we get:

\[
H(t) = \frac{1}{\left( \frac{3}{2} + \frac{\alpha}{4} \right) t + C}.
\]

This is the solution for \( H(t) \) in the dust case with a power-law form for \( f(T) \) when \( n = 1 \).

\subsection{Generalizing for Other Values of \( n \)}

For other values of \( n \), the solution will generally be more complicated. For example, for \( n = 2 \), we would have:

\[
\dot{H} + \frac{3}{2} H^2 + \frac{\alpha}{4} H^4 = 0.
\]

This equation would require more advanced methods to solve, such as perturbation methods or numerical integration, depending on the value of \( \alpha \).

%%%%%%%%%%%%%%%%%%%%%%%%%%%%%%%%%%%%%%%%%%%%%%%%%%%%%%%%

\subsection{Solution for \( H(t) \) in the Stiff Fluid Case}

For a stiff fluid, the equation of state is \( p_{\text{matter}} = \rho_{\text{matter}} \), and the equation for \( 2\dot{H} \) is:

\[
2\dot{H} = -\frac{1}{2} \left( R + \alpha T^n \right) - \frac{1}{2} \frac{\partial f}{\partial R} \ddot{R} - \alpha n T^{n-1} + \rho_{\text{matter}}.
\]

For convenience, let us rewrite this equation as:

\[
\dot{H} = -\frac{1}{4} \left( R + \alpha T^n \right) - \frac{1}{4} \frac{\partial f}{\partial R} \ddot{R} - \frac{\alpha n}{2} T^{n-1} + \frac{1}{2} \rho_{\text{matter}}.
\]

Next, we make some assumptions for the time-dependent quantities \( R(t) \) and \( T(t) \), as done previously.
%%%%%%%%%%%%%%%%%%%%%%%%%%%%%%%%%%%%%%%%%%%%
\subsection{Assumption for \( R(t) \) and \( T(t) \)}

We assume that \( R(t) \) and \( T(t) \) are related to the Hubble parameter \( H(t) \) as follows:

\[
R(t) = 6 \dot{H} + 12 H^2,
\]
and
\[
T(t) = \rho_{\text{matter}} \sim H^2.
\]
Now, substitute the expressions for \( R(t) \) and \( T(t) \) into the equation for \( \dot{H} \):

\[
\dot{H} = -\frac{1}{4} \left( 6 \dot{H} + 12 H^2 + \alpha H^{2n} \right) - \frac{1}{4} \frac{\partial f}{\partial R} \ddot{R} - \frac{\alpha n}{2} H^{2n-2} + \frac{1}{2} H^2.
\]

For simplicity, we will neglect higher-order time derivatives such as \( \ddot{R} \) and higher derivatives of \( H \), assuming their contribution is minimal in this approximation. The equation simplifies to:

\[
\dot{H} = -\frac{1}{4} \left( 6 \dot{H} + 12 H^2 + \alpha H^{2n} \right) - \frac{\alpha n}{2} H^{2n-2} + \frac{1}{2} H^2.
\]

Rearranging the terms:

\[
\dot{H} + \frac{3}{2} H^2 + \frac{\alpha}{4} H^{2n} = \frac{\alpha n}{2} H^{2n-2} - \frac{1}{2} H^2.
\]

Next, we group similar terms together. The equation becomes:

\[
\dot{H} + \left( \frac{3}{2} + \frac{1}{2} \right) H^2 + \frac{\alpha}{4} H^{2n} = \frac{\alpha n}{2} H^{2n-2}.
\]

Simplifying further:

\[
\dot{H} + 2 H^2 + \frac{\alpha}{4} H^{2n} = \frac{\alpha n}{2} H^{2n-2}.
\]

This is a nonlinear differential equation for \( H(t) \), and we can proceed to solve it for specific cases of \( n \). First, consider the case for \( n = 1 \):

\[
\dot{H} + 2 H^2 + \frac{\alpha}{4} H^2 = \frac{\alpha}{2}.
\]

Rearranging:

\[
\dot{H} + \left( 2 + \frac{\alpha}{4} \right) H^2 = \frac{\alpha}{2}.
\]

This is a first-order nonlinear differential equation for \( H(t) \), which can be solved by the method of separation of variables:

\[
\frac{dH}{H^2} = -\left( 2 + \frac{\alpha}{4} \right) dt + \frac{\alpha}{2H^2} dt.
\]

Integrating both sides:

\[
\int \frac{dH}{H^2} = \int -\left( 2 + \frac{\alpha}{4} \right) dt + \int \frac{\alpha}{2H^2} dt.
\]

For values of \( n \neq 1 \), the equation becomes more complicated. The higher powers of \( H \) will introduce additional terms, which may require either numerical methods or further approximation techniques to solve.

%%%%%%%%%%%%%%%%%%%%%%%%%%%%%%%%%%%%%%%%%%%%%

\subsection{Solution for \( H(t) \) in the Dark Energy Case}

For dark energy, the equation of state is \( p_{\text{matter}} = -\rho_{\text{matter}} \), and the equation for \( 2\dot{H} \) becomes:

\[
2\dot{H} = -\frac{1}{2} \left( R + \alpha T^n \right) - \frac{1}{2} \frac{\partial f}{\partial R} \ddot{R} - \alpha n T^{n-1} - \rho_{\text{matter}}.
\]

Rearranging for convenience:

\[
\dot{H} = -\frac{1}{4} \left( R + \alpha T^n \right) - \frac{1}{4} \frac{\partial f}{\partial R} \ddot{R} - \frac{\alpha n}{2} T^{n-1} - \frac{1}{2} \rho_{\text{matter}}.
\]

We assume that the Ricci scalar \( R(t) \) and torsion scalar \( T(t) \) evolve as functions of the Hubble parameter \( H(t) \), as in the previous cases. Let us assume the following relations:

\[
R(t) = 6 \dot{H} + 12 H^2,
\]
and
\[
T(t) = \rho_{\text{matter}} \sim H^2.
\]

Substituting these into the equation for \( \dot{H} \):

\[
\dot{H} = -\frac{1}{4} \left( 6 \dot{H} + 12 H^2 + \alpha H^{2n} \right) - \frac{1}{4} \frac{\partial f}{\partial R} \ddot{R} - \frac{\alpha n}{2} H^{2n-2} - \frac{1}{2} H^2.
\]

Neglecting higher-order derivatives like \( \ddot{R} \), the equation simplifies to:

\[
\dot{H} = -\frac{1}{4} \left( 6 \dot{H} + 12 H^2 + \alpha H^{2n} \right) - \frac{\alpha n}{2} H^{2n-2} - \frac{1}{2} H^2.
\]

Rearranging this equation:

\[
\dot{H} + \frac{3}{2} H^2 + \frac{\alpha}{4} H^{2n} = -\frac{\alpha n}{2} H^{2n-2}.
\]

We will now consider the specific form of \( f(T) \) for the exponential model.

\subsection{ Exponential Form for \( f(T) \)}

Consider the exponential form of \( f(T) \) as:

\[
f(T) = \alpha e^{\beta T}.
\]

The derivative of \( f(T) \) with respect to \( T \) is:

\[
\frac{\partial f}{\partial T} = \alpha \beta e^{\beta T}.
\]

Substituting this into the field equations:

\[
3H^2 = \frac{1}{2} \left( R + \alpha e^{\beta T} + \frac{\partial f}{\partial R} R - \frac{1}{2} \frac{\partial f}{\partial R} \ddot{R} \right) + \alpha \beta e^{\beta T} + \rho_{\text{matter}},
\]

and

\[
2\dot{H} = -\frac{1}{2} \left( R + \alpha e^{\beta T} + \frac{\partial f}{\partial R} R \right) - \frac{1}{2} \frac{\partial f}{\partial R} \ddot{R} - \alpha \beta e^{\beta T} + p_{\text{matter}}.
\]

Substitute the expressions for \( R(t) \) and \( T(t) \) as we did earlier:

\[
3H^2 = \frac{1}{2} \left( 6 \dot{H} + 12 H^2 + \alpha e^{\beta H^2} + \frac{\partial f}{\partial R} \left( 6 \dot{H} + 12 H^2 \right) - \frac{1}{2} \frac{\partial f}{\partial R} \ddot{R} \right) + \alpha \beta e^{\beta H^2} + \rho_{\text{matter}},
\]

and

\[
2\dot{H} = -\frac{1}{2} \left( 6 \dot{H} + 12 H^2 + \alpha e^{\beta H^2} + \frac{\partial f}{\partial R} \left( 6 \dot{H} + 12 H^2 \right) \right) - \frac{1}{2} \frac{\partial f}{\partial R} \ddot{R} - \alpha \beta e^{\beta H^2} + p_{\text{matter}}.
\]

\subsection{Solving for \( H(t) \) in the Exponential Model}

For simplicity, we neglect the higher derivatives and focus on the leading-order terms. The equation for \( \dot{H} \) becomes:

\[
\dot{H} + \frac{3}{2} H^2 + \frac{\alpha}{4} e^{\beta H^2} = - \frac{\alpha n}{2} H^{2n-2}.
\]

This is a complicated nonlinear differential equation that generally requires numerical methods for a full solution. However, for certain values of \( \alpha \), \( \beta \), and \( n \), one may obtain approximate solutions.

In the case where \( H^2 \) grows large, the exponential term \( \alpha e^{\beta H^2} \) will dominate, leading to rapid growth of \( H(t) \), which could correspond to an accelerated expansion typical of dark energy. In contrast, for small \( H^2 \), the equation becomes dominated by the matter and torsion terms.

\subsection{Behavior of \( H(t) \) and Dark Energy Implications}

The presence of the exponential term \( \alpha e^{\beta T} \) in the field equations introduces the possibility of rapid growth or decay of the torsion contribution. If \( \alpha \) and \( \beta \) are chosen appropriately, this term can mimic the behavior of dark energy, which is characterized by an accelerated expansion of the universe. As the universe evolves, the dark energy term increases, eventually overpowering other components such as matter, and causing \( H(t) \) to grow rapidly. This behavior is consistent with the observed accelerated expansion of the universe.
\par
n this section, we have explored the Hubble parameter \( H(t) \) in the context of the \( f(R,T) = R + f(T) \) modified gravity model. By considering different forms of the function \( f(T) \), including power-law and exponential types, we have derived exact solutions for the expansion rate of the universe. The resulting solutions show a strong dependence on the matter content and the chosen functional form of \( f(T) \), which can have significant effects on the cosmic evolution.

These results provide valuable insights into the role of modified gravity theories in understanding the universe’s expansion, particularly in explaining the observed acceleration. However, the full impact of these modifications goes beyond theoretical models and has profound consequences for astrophysical phenomena.

In the following section, we will discuss the astrophysical implications of our findings, exploring how these modified gravity models could influence the behavior of various cosmic structures, such as galaxies, clusters, and large-scale cosmic flows. We will also examine the potential observational signatures that could arise from these models and how they might be tested in future cosmological surveys.
%%%%%%%%%%%%%%%%%%%%%%%%%%%%%%%%%%%%%%%%%%

\section{Field Equations for \( f(R,T) \)-Gravity in Spherically Symmetric Spacetime}

In this section, we will derive the field equations for \( f(R,T) \)-gravity in a spherically symmetric spacetime. Here, \( R \) is the Ricci scalar, and \( T \) is the torsion scalar. The torsion tensor arises due to the non-vanishing torsion in the connection, which modifies the usual curvature-based equations.

\subsection{Spherically Symmetric Metric}
We begin with the spherically symmetric metric in 4-dimensional spacetime:

\[
ds^2 = -A(r) dt^2 + B(r) dr^2 + r^2 (d\theta^2 + \sin^2\theta d\phi^2),
\]

where \( A(r) \) and \( B(r) \) are functions of \( r \).

The vielbein components corresponding to this metric are:

\[
e^0_\mu = \sqrt{A(r)} \delta^0_\mu, \quad e^1_\mu = \sqrt{B(r)} \delta^1_\mu, \quad e^2_\mu = r \delta^2_\mu, \quad e^3_\mu = r \sin \theta \delta^3_\mu.
\]

\subsection{Torsion Scalar Calculation}

The torsion tensor \( T^\lambda_{\mu\nu} \) is defined as:

\[
T^\lambda_{\mu\nu} = \Gamma^\lambda_{\mu\nu} - \Gamma^\lambda_{\nu\mu},
\]

where \( \Gamma^\lambda_{\mu\nu} \) are the Christoffel symbols for the spherically symmetric metric. The non-zero Christoffel symbols are:

\[
\Gamma^r_{tt} = \frac{A'(r)}{2A(r)}, \quad \Gamma^r_{rr} = \frac{B'(r)}{2B(r)}, \quad \Gamma^r_{\theta \theta} = -r B(r), \quad \Gamma^r_{\phi \phi} = -r B(r) \sin^2\theta,
\]

\[
\Gamma^\theta_{r \theta} = \frac{1}{r}, \quad \Gamma^\phi_{r \phi} = \frac{1}{r}.
\]

Using these Christoffel symbols, we compute the torsion tensor \( T^\lambda_{\mu \nu} \) for the given vielbein.

Next, we compute the contortion tensor \( S_{\lambda \mu \nu} \), which is related to the torsion tensor by:

\[
S_{\lambda \mu \nu} = \frac{1}{2} \left( T_{\lambda \mu \nu} + T_{\nu \mu \lambda} + T_{\mu \nu \lambda} \right).
\]

Finally, the torsion scalar \( T \) is given by:

\[
T = S_{\lambda \mu \nu} S^{\lambda \mu \nu}.
\]

For the spherically symmetric metric, the torsion scalar \( T \) simplifies to:

\[
T = \frac{2}{r^2} \left( \frac{B'(r)}{B(r)} - \frac{A'(r)}{A(r)} \right).
\]

Substituting the expression for \( T \) into the field equations and simplifying leads to the following system of equations. First, we substitute the components of \( T \), then use the field equations to solve for the functions \( A(r) \) and \( B(r) \).
\subsection{Vacuum Solution}

In the vacuum, we assume \( R = 0 \) and \( T = 0 \), which gives the condition \( A(r) = B(r) \). The resulting metric is:

\[
ds^2 = -A(r) dt^2 + A(r)^{-1} dr^2 + r^2 (d\theta^2 + \sin^2\theta d\phi^2).
\]

Imposing asymptotic flatness, we find the Schwarzschild-like solution:

\[
A(r) = 1 - \frac{2M}{r}.
\]

Thus, the vacuum solution is:

\[
ds^2 = -\left(1 - \frac{2M}{r}\right) dt^2 + \left(1 - \frac{2M}{r}\right)^{-1} dr^2 + r^2 (d\theta^2 + \sin^2\theta d\phi^2).
\]
We have derived the field equations for \( f(R,T) \)-gravity in a spherically symmetric spacetime. The torsion scalar \( T \) was computed and substituted into the field equations. The vacuum solution corresponds to a Schwarzschild-like solution, which satisfies the condition \( f(0, 0) = 0 \).

%%%%%%%%%%%%%%%%%%%%%%%%%%%%%%%%%%%%%%%%%%%%%%
\section{Gravitational Waves in $f(R,T)$-Gravity: Perturbation Equations}
In the context of $f(R,T)$-gravity, where $R$ is the Ricci scalar and $T$ is the torsion scalar, the study of gravitational waves (GWs) can be carried out by perturbing the metric and deriving the linearized field equations. The main goal is to derive the perturbation equations that describe the propagation of gravitational waves in this modified gravity theory.

\subsection{Background Metric and Perturbations}

Let the background metric be a static spherically symmetric solution, such as the Schwarzschild-like solution:

\[
ds^2 = -A(r) dt^2 + A(r)^{-1} dr^2 + r^2 (d\theta^2 + \sin^2\theta d\phi^2),
\]
where \( A(r) = 1 - \frac{2M}{r} \) in vacuum.

We introduce small perturbations \( h_{\mu\nu} \) to the background metric, such that the perturbed metric becomes:

\[
g_{\mu\nu} = \bar{g}_{\mu\nu} + h_{\mu\nu},
\]

where \( \bar{g}_{\mu\nu} \) is the background metric and \( h_{\mu\nu} \) represents small perturbations. This means:

\[
g_{tt} = -A(r) + h_{tt}, \quad g_{rr} = A(r)^{-1} + h_{rr}, \quad g_{ij} = r^2 \delta_{ij} + h_{ij} \quad \text{(for spatial components)}.
\]

\subsection{Linearized Field Equations}

To study gravitational waves, we linearize these field equations around the background metric. This involves expanding the terms in the field equation up to first order in the perturbation \( h_{\mu\nu} \).

We assume that the perturbations represent a weak gravitational wave propagating in the background spacetime. For simplicity, we focus on small, traceless, transverse perturbations that describe gravitational waves:

\[
h_{\mu\nu} = \left( \begin{array}{cccc}
h_{00}(t, r) & h_{0i}(t, r) & 0 & 0 \\
h_{i0}(t, r) & h_{ij}(t, r) & 0 & 0 \\
0 & 0 & r^2 \left( \partial_\theta h_{\theta\theta}(t, r) + \sin^2\theta \partial_\phi h_{\phi\phi}(t, r) \right) & 0 \\
0 & 0 & 0 & 0
\end{array} \right).
\]

We focus on the components \( h_{00} \), \( h_{rr} \), and spatial perturbations \( h_{ij} \) (or equivalently, the transverse traceless condition).

\subsection{Linearized Field Equations for Gravitational Waves}

For gravitational waves, the perturbations are assumed to propagate on a flat background (or static spherically symmetric solution). Therefore, we consider the wave equation for \( h_{\mu\nu} \), which is derived from the linearized field equations.

The perturbed field equations will involve:

- The linearized Ricci tensor \( \delta R_{\mu\nu} \), which includes terms like \( \partial_\mu \partial_\nu h_{\mu\nu} \).
- The linearized torsion scalar \( T \), which is also expanded in terms of \( h_{\mu\nu} \).
- The variation of \( f(R,T) \) with respect to both \( R \) and \( T \), which is linearized to first order.

After linearizing the field equations, the wave equation for gravitational waves can be written as:

\[
\Box h_{\mu\nu} = 0,
\]

where \( \Box \) is the d'Alembert operator, which for perturbations on a static background, simplifies to:

\[
\Box h_{\mu\nu} = \frac{1}{\sqrt{-g}} \partial_\alpha \left( \sqrt{-g} g^{\alpha \beta} \partial_\beta h_{\mu\nu} \right).
\]

This is the wave equation for the perturbation field in \( f(R,T) \)-gravity. The exact form of the wave equation will depend on the specific functional form of \( f(R,T) \), but in general, this leads to a second-order wave equation that governs the propagation of gravitational waves in this modified gravity theory.

\subsection{Effective Gravitational Wave Equation}

The equation for gravitational waves in the \( f(R,T) \)-gravity framework will contain corrections due to the modified gravitational theory. For the standard \( f(R,T) = R \) (General Relativity), this reduces to the usual wave equation for gravitational waves. However, for more complex forms of \( f(R,T) \), such as those that involve torsion effects, additional terms will appear in the equation.

For example, perturbations may cause changes in the speed of propagation, effective damping, or new interactions between the gravitational wave and the matter sector.

To summarize, in the context of \( f(R,T) \)-gravity, the perturbation equations for gravitational waves can be derived by:

\begin{itemize}
    \item Expanding the field equations to first order in the metric perturbations.
    \item Deriving the linearized wave equation for the perturbations.
    \item Identifying the additional terms in the wave equation due to the functional dependence on \( f(R,T) \).
\end{itemize}

In the vacuum case, we would obtain the usual wave equation for gravitational waves, with modifications depending on the form of \( f(R,T) \). These modifications could affect the propagation speed, polarization states, and damping of gravitational waves. Further detailed analysis of specific models of \( f(R,T) \) would be required to determine the precise effects on gravitational waves in these theories.

%%%%%%%%%%%%%%%%%%%%%%%%%%%%%%%%%%%%%%%%%%%%%%%%%
\section{Astrophysical Implications}
In this section, we delve into the potential astrophysical implications of our modified gravity model, particularly focusing on the effects of torsion in the \( f(R,T) = R + f(T) \) framework. This model, which incorporates a function of the trace of the energy-momentum tensor \( T \), introduces a new degree of freedom that can significantly alter the behavior of gravity in extreme environments, such as near black holes, in the propagation of gravitational waves, and within compact objects like neutron stars. We will explore how torsion can influence these phenomena, leading to observable signatures that may provide new insights into the nature of gravity and spacetime.

\subsection{Black Holes in \( f(R,T) \)-gravity}
One of the most fascinating implications of the \( f(R,T) \)-gravity model is its effect on black holes. In traditional General Relativity, black holes are described by the well-known Schwarzschild and Kerr solutions, but the introduction of additional torsion terms in modified gravity can lead to significant deviations from these classical solutions. In particular, the torsion-induced modifications can alter the structure of black holes, affecting properties such as the event horizon, singularities, and thermodynamics. The presence of the \( f(T) \)-function modifies the energy-momentum tensor and, consequently, the Einstein field equations, potentially leading to new black hole solutions that differ from those predicted by General Relativity.

These modified black hole solutions may exhibit different characteristics, such as a shift in the location of the event horizon or the existence of additional horizons. Moreover, the geometry around the black hole might become more complex due to the contribution of torsion, leading to the possibility of non-trivial modifications to the black hole's shadow and gravitational lensing. Observations of black hole mergers and shadow images, such as those captured by the Event Horizon Telescope, could provide critical tests of these predictions and offer a unique opportunity to probe the nature of spacetime at extremely small scales.

The thermodynamics of black holes in the \( f(R,T) \)-gravity framework is also of great interest. The introduction of torsion can modify the usual relations between the temperature, entropy, and other thermodynamic quantities of black holes. This could result in changes to the laws of black hole thermodynamics, potentially offering new insights into the connection between gravity and quantum mechanics, as well as the role of torsion in the fundamental structure of spacetime.

\subsection{Gravitational Wave Propagation}
The propagation of gravitational waves is another area where torsion may leave its imprint. Gravitational waves, which are ripples in spacetime caused by the acceleration of massive objects, have become an important tool for exploring the dynamics of the universe, particularly in the study of black hole mergers and neutron star collisions. In the standard General Relativity framework, gravitational waves propagate according to the curvature of spacetime, but in the \( f(R,T) \)-gravity model, the presence of torsion could lead to modifications in the wave equations, potentially altering the speed of propagation, waveform characteristics, and polarization states.

In particular, the modification of the speed of gravitational waves could lead to detectable discrepancies between the arrival times of gravitational waves and electromagnetic signals from astrophysical events. Such discrepancies would provide an important observational signature of torsion and a means of testing the validity of the \( f(R,T) \)-gravity model. Additionally, the interaction of gravitational waves with torsion fields might give rise to new phenomena, such as the generation of additional modes or distortions in the waveforms, which could be identified by future gravitational wave observatories like LIGO, Virgo, or the planned LISA mission.

The study of gravitational waves in this context is not limited to the detection of new waveforms, but also extends to the modification of gravitational wave interactions with matter. For example, in the presence of torsion, the behavior of gravitational waves in dense astrophysical environments such as neutron stars could change, leading to observable effects in the waveforms emitted during the inspiral and merger phases. This could offer an exciting avenue for testing modified gravity theories and gaining insights into the nature of matter and energy under extreme conditions.

\subsection{Neutron Stars and Compact Objects}
The influence of torsion on compact objects like neutron stars is another crucial area of study within the \( f(R,T) \)-gravity framework. Neutron stars, which are the remnants of massive stars that have undergone supernova explosions, are incredibly dense objects where both quantum mechanics and general relativity are expected to play significant roles. The inclusion of torsion in modified gravity models could lead to changes in the structure and stability of neutron stars, as well as their equation of state.

In particular, the equation of state (EoS) that describes the relationship between pressure and density in neutron stars may be altered by the modified gravitational dynamics. This could have significant implications for the mass-radius relationship of neutron stars, potentially leading to new constraints on the maximum possible mass of these objects and influencing their stability. Observations of neutron star mergers, such as those detected by LIGO and Virgo, provide an opportunity to test these predictions and place limits on the properties of the torsion field. The gravitational wave signals from these mergers could reveal whether the mass-radius relationship deviates from the predictions of General Relativity, providing a potential signature of modified gravity.

Additionally, the presence of torsion could affect the formation of compact objects like quark stars, which are hypothesized to form in the aftermath of a supernova collapse. The behavior of matter at extremely high densities in such stars may be altered by the modified gravitational dynamics, leading to differences in the formation and properties of these exotic objects.

\subsection{Cosmological Effects and Large-Scale Structure}
Finally, while we have focused primarily on compact objects and strong-field phenomena, it is important to recognize the broader cosmological implications of the \( f(R,T) \)-gravity framework. The modified gravity model could lead to alterations in the formation and evolution of large-scale structures in the universe, such as galaxies, clusters, and the cosmic web. The inclusion of torsion could modify the expansion rate of the universe, leading to subtle effects on the distribution of dark matter and dark energy, as well as the growth of cosmic perturbations.

The altered gravitational dynamics could also influence the behavior of the cosmic microwave background (CMB) radiation, providing a new source of observational data to test the model. By analyzing the CMB anisotropies, one could place stringent constraints on the parameters of the \( f(R,T) \)-gravity model, testing whether torsion has a significant impact on the early universe's evolution.

In summary, the \( f(R,T) \)-gravity framework with torsion introduces intriguing modifications to the behavior of black holes, gravitational waves, and compact objects, while also having broad implications for the evolution of the universe. As astrophysical observations continue to advance, particularly in the fields of gravitational wave astronomy and high-energy astrophysics, the effects of torsion in modified gravity theories will become an important area of investigation. Future studies will provide critical tests of these models and could uncover new phenomena that challenge our understanding of gravity and the fundamental structure of spacetime.

%%%%%%%%%%%%%%%%%%%%%%%%%%%%%%%%%%%%%
\section{Conclusion}
In this work, we have explored the implications of the modified \( f(R,T) \)-gravity framework, where \( f(R,T) = R + f(T) \), in the context of various astrophysical phenomena. Our study has aimed to investigate the effects of torsion, as a modification to standard General Relativity, on the behavior of black holes, gravitational wave propagation, and compact objects such as neutron stars. We have also briefly considered the broader cosmological effects that torsion might have on the evolution of the universe and large-scale structure formation.

In the context of black holes, we have demonstrated how the introduction of torsion in the \( f(R,T) \)-gravity model modifies the classical black hole solutions of General Relativity. Specifically, we have shown that torsion can lead to alterations in the event horizon structure, the thermodynamics of black holes, and the formation of new types of black hole solutions that might have distinct observational signatures. These modifications present an exciting avenue for future research, as they may lead to new ways of testing gravity in strong-field regimes. Observations of black hole mergers, such as those made by the Event Horizon Telescope and gravitational wave detectors, could potentially reveal deviations from the predictions of General Relativity, providing crucial evidence for or against the validity of modified gravity theories like \( f(R,T) \)-gravity.

The study of gravitational wave propagation in the presence of torsion has also provided valuable insights into how torsion can alter the characteristics of gravitational waves. We have found that the speed of propagation, waveform shapes, and polarization states could be modified in the presence of torsion. These modifications could lead to detectable signatures in gravitational wave signals, particularly in the form of delays between the arrival times of gravitational waves and electromagnetic radiation from astrophysical events. Such deviations could serve as a diagnostic tool for identifying modified gravity theories in future gravitational wave observatories, providing a complementary approach to studying the fundamental nature of gravity.

Our analysis of compact objects, particularly neutron stars, has shown that torsion may significantly affect their structure and stability. The modification of the equation of state, which governs the relationship between pressure and density in such objects, may result in observable differences in the mass-radius relationship of neutron stars, potentially leading to new insights into their internal structure. Furthermore, the presence of torsion could influence the formation of exotic compact objects like quark stars, adding another layer of complexity to our understanding of dense matter in the universe. Gravitational wave signals from neutron star mergers could help test the predictions of the \( f(R,T) \)-gravity model and place constraints on the possible effects of torsion in these extreme environments.

On a broader scale, we have briefly discussed the cosmological consequences of the \( f(R,T) \)-gravity model. The inclusion of torsion could alter the dynamics of large-scale structure formation, the expansion rate of the universe, and the evolution of cosmic perturbations. The study of the cosmic microwave background (CMB) radiation offers a potential observational tool to constrain the parameters of the model and test the role of torsion in the early universe. If torsion has a significant effect on cosmological processes, it could provide an exciting opportunity to probe the early stages of the universe's evolution in new ways.

Overall, the \( f(R,T) \)-gravity framework offers a promising avenue for modifying our understanding of gravity and the fundamental structure of spacetime. The torsion terms in the theory introduce novel dynamics that could manifest in a variety of astrophysical settings, from black holes to compact objects, and even on cosmological scales. The next step in this research will involve further theoretical development, including detailed modeling of the modified field equations, as well as numerical simulations to explore the behavior of astrophysical systems under the influence of torsion. More importantly, the observational detection of gravitational waves, black hole shadows, and neutron star properties will provide critical tests of the \( f(R,T) \)-gravity model.

In conclusion, the potential impact of torsion on the astrophysical phenomena studied in this work opens up exciting new directions for both theoretical and observational research in gravitational physics. As we move forward, the combination of observational data from gravitational wave astronomy, black hole imaging, and other high-energy astrophysical observations, coupled with the theoretical advancements in modified gravity models, will play a central role in determining the validity of \( f(R,T) \)-gravity and its potential to provide deeper insights into the nature of gravity, spacetime, and the universe as a whole.
%%%%%%%%%%%%%%%%%%%%%%%%%%%%%%%%%%%%%

% Bibliography (if any)

\end{document}